# Link Graph Analysis for Adult Images Classification

Evgeny Kharitonov
Moscow Institute of Physics and Technology, Yandex LLC
119021, 16 Lev Tolstoy st.,
Moscow
*kharitonov@yandex-team.ru*

Anton Slesarev
Moscow Institute of Physics and Technology, Yandex LLC
119021, 16 Lev Tolstoy st.,
Moscow
*slesarev@yandex-team.ru*

Ilya Muchnik
Rutgers University, Yandex LLC
119021, 16 Lev Tolstoy st.,
Moscow
*muchnikilya@yahoo.com*

Fedor Romanenko
Yandex LLC
119021, 16 Lev Tolstoy st.,
Moscow
*fedor@yandex-team.ru*

Dmitry Belyaev
Yandex LLC
119021, 16 Lev Tolstoy st.,
Moscow
*dvbelyaev@yandex-team.ru*

Dmitry Kotlyarov
Yandex LLC
119021, 16 Lev Tolstoy st.,
Moscow
*dimko@yandex-team.ru*

## ABSTRACT

In order to protect an image search engine's users from undesirable results adult images' classifier should be built. The information about links from websites to images is employed to create such a classifier. These links are represented as a bipartite website-image graph. Each vertex is equipped with scores of adultness and decentness. The scores for image vertexes are initialized with zero, those for website vertexes are initialized according to a text-based website classifier. An iterative algorithm that propagates scores within a website-image graph is described. The scores obtained are used to classify images by choosing an appropriate threshold. The experiments on Internet-scale data have shown that the algorithm under consideration increases classification recall by 17% in comparison with a simple algorithm which classifies an image as adult if it is connected with at least one adult site (at the same precision level).

## 1. INTRODUCTION

There are two kinds of approaches which can be used to detect adult images, i.e. text-based and image-based. The text-based approach detects adult webpages using their text content and propagates this information to the linked images and pages. On the other hand, image-based approach uses the features contained in the image itself such as face presence, skin-color features, connected components, etc. The problem of adult webpages detection problem is a special case of the automatic text classification problem.

There are many papers in this field which differ in feature extraction and selection methods, machine learning algorithms used and some other details. Sebastiani made a survey [1] of the main approaches to text classification and categorization. Text-based approaches are full of significant limitations. Texts on many webpages do not correspond to their image contents. For example, a webmaster can mislead search engines trying to show adult images in response to popular decent queries to attract a user's attention. Some pages lack texts for good classification. Usually a special dictionary for text classifying is needed. Because of limitations on size of a training set and variety of words used on adult webpages it can be difficult to make such a dictionary. Another aspect is that many words and phrases may be obscene or decent depending on the context. Thus a context analysis is a hard task.

Text-based classifiers could be applied to bigger structures such as websites. In work [2] website adultness is defined by the number of adult webpages on this website. The use of websites instead of webpages decreases classification noise.

The problem of adult image classification using image content is widely investigated in literature. For example, Rowley et al. [3] used image features and proposed algorithm that reach 90% precision and 50% recall.

There are papers which use a combination of text and images features, for example [4].

Further in the paper we will firstly introduce some related work and then present our method. After that we will provide some experimental results on a real dataset. Finally, we will point out some possible directions of our future work.

## 2. Related Work

Considering the adult image classification problem based on an adult websites classifier the notion of a bipartite image-website graph appears. An edge in this graph indicates whether the website and the images are linked. The problem we address is how to classify adult images using the information about websites adultness. The simplest classifier supposes an image to be adult if it appears on at least one adult website.

The idea to analyze internet data by exploration event moving on a huge graph appeared about 10 years ago. Most popular algorithms of this type are PageRank [5] and HITS [6]. Since the paper [5] was published the internet data analysis using iterational process on some graph (often such processes describe Markov random walks) has become widespread for data classification.

Castillo et al. [7] used this idea to detect spam vertices in the graph of webpages.

Li et al. [8] used click graphs to improve query intent classifiers. A click graph is a bipartite graph representing click-through data. The edges therein connect queries and URLs and are weighed by the associated click counts. The authors manually labeled a small set of seed queries in a click graph and iteratively propagated the label information to the other queries until a global equilibrium state was reached.

The authors of [9] used modified HITS algorithm to detect paid links.

Szummer and Craswell [10] classified webpages by analyzing a bipartite query-URL graph. The authors labeled a small set of webpages and proposed a procedure used to classify all webpages using random walks on this graph.

Deng et al. (Deng, Lyu, and King 2009) proposed a general CO-HITS algorithm. The authors applied their algorithm to a query suggestion problem.

In the present work we describe an iterative procedure of website adultness propagation across a bipartite image-website graph. This procedure is used as part of the adult image filtering system in the image search service of search engine Yandex.

## 3. Data

The goal of the research is to classify images as adult or decent provided the text-based site classification. The algorithm uses undirected bipartite graph $G = (V; E) = (S; I; E)$ which represents the links between websites and images in the Internet.

The vertexes of the partite $S$ represent the websites named by their URLs. All sites' URLs are truncated to their second domain except for several hostings (there are about 20 hostings in the exception list). For instance, we do not distinguish 1.regularhost.com, 2.regularhost.com and regularhost.com and treat them as a single vertex, but user1.livejournal.com and user2.livejournal.com are mapped to different vertexes.

Partite $I$ consists of indexed images grouped into clusters by their visual similarity using Dmitry Mikhalev's algorithm[1], so every vertex $i \in I$ represents a cluster of similar images. Clusterization is used in order to group resized and compressed copies of an image. Such clusterization reduces the graph size and decreases sparseness of the graph's incidence matrix.

$V = S \cup I$ is a set if vertexes of $G$.

A vertex from the first partite is connected with a vertex from the second one if there is a link (html tags "img" or "a href" ) from a page on a website to at least one image in a corresponding image cluster. It should be noticed that no distinction is made between cases with many and a few links between a site and an image cluster. We also suppose that all sites have been already classified as adult or not. This initial site classification might have errors.

In this paper we use the results of site classification produced by the text-based site classifier described in [2].

## 4. Data Model and Algorithm

We denote the image-site graph as $G(V; E) = G(S^+ \cup S^-; I; E)$, where $S^+$ stands for a set of adult websites, $S^-$ stands for the rest of websites (regarded decent). The image set (the set of image clusters) is referred to as $I$. Our goal is to classify images $i \in I$ into two classes.

With the graph $G$ we associate a binary adjacency matrix $W(G)$. $w_{ij} = 1$ if there is an edge between vertexes $i$ and $j$ and $w_{ij} = 0$ otherwise.

Let us consider a matrix $Y(V)$ which characterizes initial label distribution on the set of vertexes. Its size is $|V| \times 2$ with rows enumerating graph vertexes and columns corresponding to class labels. $y_{i1} = 1, y_{i2} = 0$ if vertex $i$ is labeled as adult and $y_{i1} = 0, y_{i2} = 1$, if $i$ is labeled as decent. $y_{i1} = y_{i2} = 0$, if vertex $i$ has no label (this holds for all images $i \in I$).

We are looking for a $|V| \times 2$ non-negative matrix $F(V)$. Vector $F_i = (F_{i1}, F_{i2})$ is interpreted as $i$ vertex's class score, $F_{i1}$ and $F_{i2}$ are adultness and decentness, respectively.

$Y_i$ is taken as initial value of $F_i^{(0)}$: $F_i^{(0)} = Y_i, i \in V$

The problem of computing $F$ is an optimization problem. For every edge $(i, j)$ in the graph we are going to find a vector pair $(F_i; F_j)$ such that $F_i$ is close to $F_j$.

The intuition behind these criteria is that for every pair of adjacent vertexes we want their scores of label intensity to be as close as possible. We choose the following function as a criterion:

---

[1] Unfortunately, the algorithm's details have not been published. The average size of a cluster is 1.7.

$$\Phi = \frac{1}{2}\sum_{i,j\in V} W_{ij}\left[\frac{F_i}{\sqrt{D_{ii}}} - \frac{F_j}{\sqrt{D_{jj}}}\right]^2 + \mu\sum_{i\in V}(F_i - Y_i)^2$$

Here $D$ denotes a diagonal matrix with elements $D_{ii} = \sum_{j\in V} W_{ij}$ and $\mu$ is the algorithm's parameter.

We adopt this function from paper [8] where the authors consider the problem of label propagation on a set of search queries.

The first term denotes our wish for $F_i$ and $F_j$ to be close. It contains three weight multiples $D_{ii}$, $D_{jj}$ and $W_{ij}$. $W_{ij}$ shows that we summarize discrepancy only over the graph edges. $D_{ii}$ and $D_{jj}$ are used to decrease class scores for vertexes with a lot of incident edges.

The second term is used to limit deviation of resultant weights from initial weights. Coefficient $\mu$ regularizes the bounds of this deviation.

The optimization can be derived analytically:

$$F^* = \arg\min \Phi = (1-\alpha)(E - \alpha A)^{-1} Y,$$

$$A = D^{-1/2} W D^{-1/2},$$

$$\alpha = \frac{1}{1+\mu},$$

$$E = diag(1,...,1)$$

But there is a convenient iterative procedure to find $F^*$ that exploits a sparse structure of $W$ and allows us to avoid expensive calculation of the large inverse matrix:

$$F^{(0)} = Y,$$

$$F^{(n+1)} = \alpha A F^{(n)} + (1-\alpha) Y$$

It can be shown ([8]) that this procedure converges to the analytical solution.

For each vertex $i \in V$ we get a vector $(F_{i1}^*, F_{i2}^*)$ as a result of calculation $F^* = \arg\min \Phi$. In the next step we sort all the vertexes in the decreasing order of $\frac{F_{i1}^*}{F_{i2}^* + \varepsilon}$. After that we label $\lfloor |V| \cdot k \rfloor$ top vertexes with maximum value of ratio $\frac{F_{i1}^*}{F_{i2}^* + \varepsilon}$ as adult. Here $\varepsilon$ is some small constant and $k$ is used to adjust the tradeoff between recall and precision.

## 5. Experiments

In our experiments we use the image-site graph and the results of automatic website classification built on 20.05.2009. No additional information is used. We remove the sites without images, because our goal is to classify images. We also remove small images[2]. The resultant graph consists of $1.6\times10^6$ sites, $444\times10^6$ image groups and $735\times10^6$ edges. The average number of images connected with a site is 460.

---

[2] Images with area less than $100\times50$. Most of them are very uninformative: icons, avatars, buttons, etc.

To the estimate quality of the image classifier we use an additional dataset. All images in the dataset were labeled manually as adult or not. The dataset consists of 9475 random images collected from the internet. 117 of them are adult. We have performed 42 experiments depending on 3 parameters. These parameters are:

- the number of iterations $(n)$,
- the parameter which regularizes discrepancy of $F^*$ from $Y$ $(\alpha)$,
- the size of adult images class $(k)$

$\varepsilon$ is a constant in all experiments. Its value is 0.001.

The following scheme is used to choose the parameters' value. First of all we choose $\alpha = 0.5$ from the problem's specifics (after that $\alpha$ was varied in 0.2…0.8). We also take $k = 0.04$ as reasonable according to the problem specifics. We have performed experiments with these parameters with $n$ varied from 1 up to 30 (Fig. 1, Fig. 2). After determining the best value of $n$ we vary $\alpha \in \{0.2, 0.4, 0.6, 0.8\}$ and $k \in \{0.01, 0.02, 0.03, 0.04, 0.06, 0.08, 0.10, 0.12\}$ (32 combinations total) with fixed $n$ (Fig. 3).

To estimate the classification quality we use two metrics, i.e. precision and recall. Let us consider a task of classifying instances $S$ into classes $\{+1, -1\}$. Experts label $S_{+1}{}^{e} \subset S$ as instances of class $+1$ and $S_{-1}{}^{e} \subset S$ as instances of class $-1$. The classifier splits set $S$ into two sets of instances $S_{+1}{}^{c}$ and $S_{-1}{}^{c}$ classified as +1 and -1, correspondingly. Then

$$precision = \frac{|S_{+1}{}^{e} \cap S_{+1}{}^{c}|}{|S_{+1}{}^{c}|},$$

$$recall = \frac{|S_{+1}{}^{e} \cap S_{+1}{}^{c}|}{|S_{+1}{}^{e}|}.$$

## 6. Discussion

At first it should be noticed that the metrics converged (Fig. 1, 2) after a few iterations were performed. Five iterations seem to be enough for all future experiments. Moreover this fact implies some practical consequences, since the iterations number affects computational complexity.

As it can be seen from Fig. 3 the variation of $\alpha$ with fixed $k$ produced a very weak effect on both recall and precision while the variation of parameter $k$ allows us to change recall and precision within a wide range.

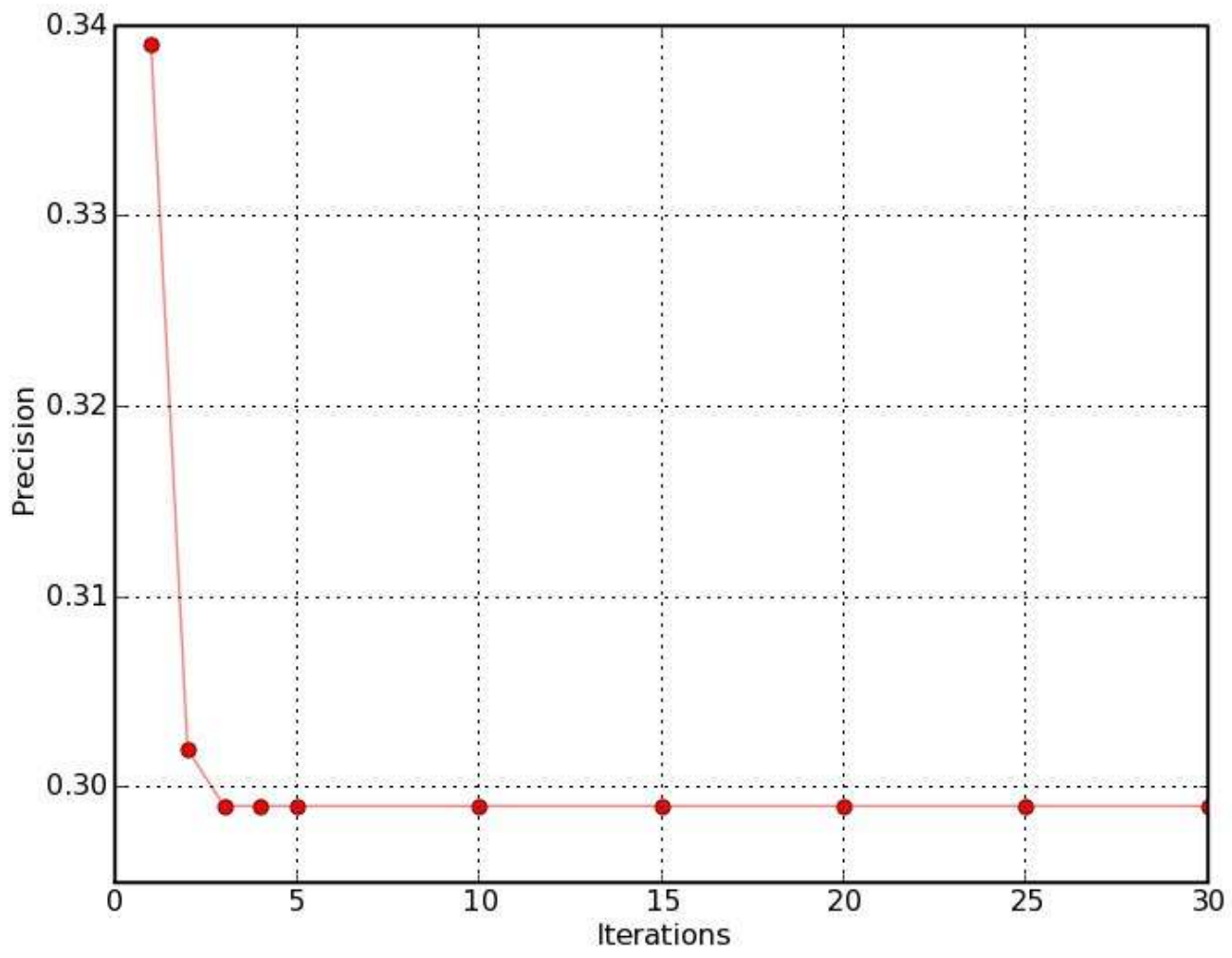

**Figure 1. Precision vs. Iterations**

These experiments allow us to compare the algorithm under consideration with the 'naive' rule that classifies all the images with at least one link from the adult site as adult themselves and decent otherwise. The algorithm outperforms a simple (referred as 'baseline' on Fig. 3) rule which classifies an image as adult if it is connected with at least one adult site by 17% in recall without changes in precision. The algorithm's recall is very important due to task specifics.

## 7. Conclusions and Future Work

We have presented a new adult image classification algorithm that makes use of image-website link structure provided prior site classification results. The following intuition is behind it:

- if a lot of suspicious sites have a picture on their pages, then the picture is a suspicious one,
- if a site contains a lot of suspicious images, then it is suspicious itself.

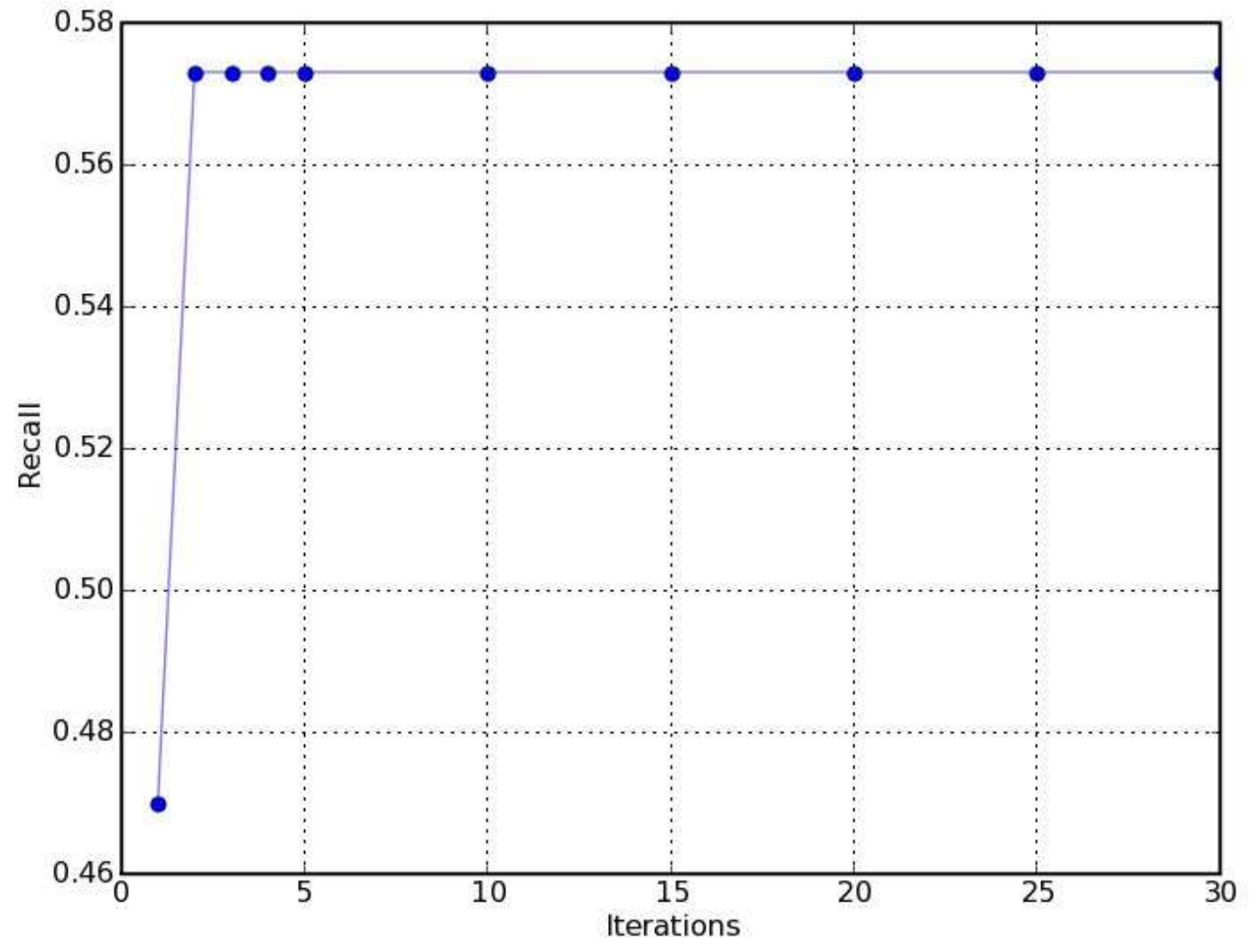

Figure 2. Recall vs. Iterations

In addition we cluster images. This makes the image-site graph less sparse, reduces the vertexes number and provides more information about a single image in the internet. We believe clustering to be important in our task because many adult sites copy adult images from each other. The novelty of the approach is the use of the image-site graph in order to solve the classification problem. Our experiments have shown that the system scales well and performs reliably as part of a production system.

The described procedure can be also applied to graph classification tasks of different nature.

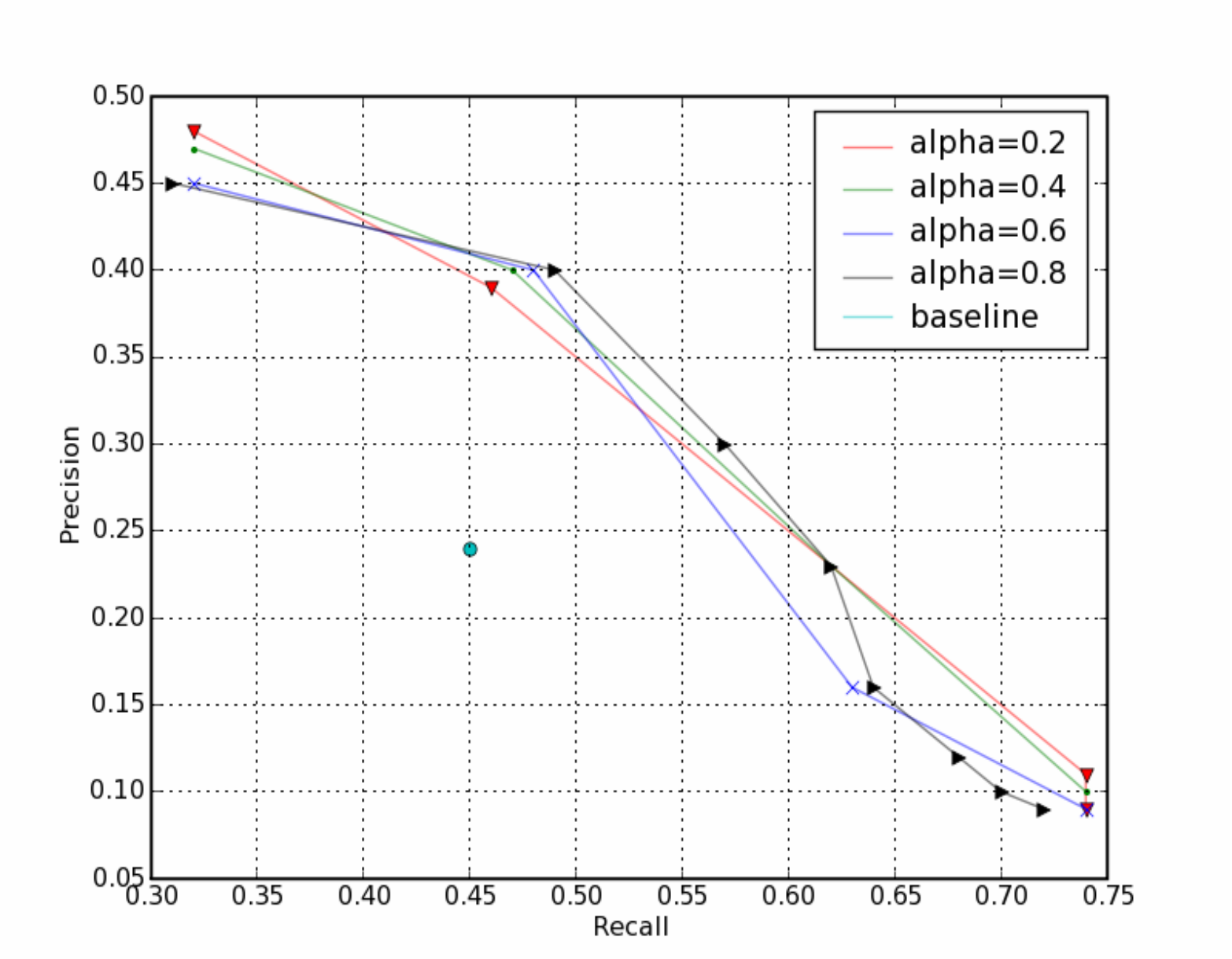

Figure 3. Precision vs. Recall

For instance, the procedure can be used in order to detect spam sites using the internet link structure or to detect commercial images using the website-image links.

For future work our short term focus is to use textual caption-based image classifiers. It would provide us with labels for the image partite of the graph that promises new resources for the algorithm improvement. We are going to investigate how using of individual pages instead of sites affects algorithm's performance.